\documentclass[aps,prl,twocolumn,showpacs]{revtex4}
\usepackage[T1]{fontenc}
\usepackage[latin1]{inputenc}
\usepackage{graphicx}
\usepackage{hyperref}
\usepackage{amsmath}

\begin{document}

\title{$q$-Breathers and thermalization in acoustic chains with arbitrary nonlinearity index}

\author{M.\,V.\,Ivanchenko}

\affiliation{Department of Applied Mathematics, University of Leeds, LS2 9JT, Leeds, United Kingdom}

\begin{abstract}
Nonlinearity shapes lattice dynamics affecting vibrational spectrum, transport and thermalization phenomena. Beside breathers and solitons one finds the third fundamental class of nonlinear modes -- $q$-breathers -- periodic orbits in nonlinear lattices, exponentially localized in the reciprocal mode space. To date, the studies of $q$-breathers have been confined to the cubic and quartic nonlinearity in the interaction potential. In this paper we study the case of arbitrary nonlinearity index $\gamma$ in an acoustic chain. We uncover qualitative difference in the scaling of delocalization and stability thresholds of $q$-breathers with the system size: there exists a critical index $\gamma^*=6$, below which both thresholds (in nonlinearity strength)  tend to zero, and diverge when above. We also demonstrate that this critical index value is decisive for the presence or absense of thermalization. For a generic interaction potential the mode space localized dynamics is determined only by the three lowest order nonlinear terms in the power series expansion.
\end{abstract}


\maketitle

Nonlinearity is a key player in a number of fundamental dynamical and statistical physical phenomena including thermal
conductivity, wave excitation and propagation, electron and phonon scattering. Its impact is often counterintuitive. E.~Fermi, J.~Pasta and S.~Ulam (FPU) hypothesized that it underlies thermalization in crystals and ran the celebrated computational experiment with oscillatory chains with nonlinearity in coupling \cite{fpu}.
Paradoxically, the low wave number $q$ excitations did not spread over the spectrum, staying localized in several neighbor modes. More than that, it showed almost exact recurrencies of the energy to the initial mode at large times. 

At present we know that this scenario is generic and holds in a certain parameter domain. There exist weak and strong stochasticity (chaos) thresholds in the energy/nonlinearity strength: above the former oscillations become chaotic, but stay localized in the mode space; only above the latter the oscillations delocalize rapidly and thermalization occurs \cite{Ford, chaosfpu, Izrailev}.
 
Despite certain achievements in the continuum-limit approximation with integrable soliton dynamics \cite{Zabusky} and analytic and numeric estimates of the strong chaos threshold \cite{Izrailev_Chirikov,deLuca, Shepel,italian,kantz,lcmcsmpegdc97}, only the recently developed theory of $q$-breathers (QB) has explained all main ingredients of the FPU problem \cite{we_qb}. 

QBs are exact periodic solutions, continuations of the linear modes in the nonlinear regime, exponentially localized in the $q$-space and stable in the suitable parameter range. FPU trajectories were shown to be small perturbations near stable QBs, hence the absence of thermalization and recurrencies. The instability of QBs was found to coincide with the weak stochasticity threshold, and delocalization -- with the strong stochasticity one and thermalization.


However, there is a principal question that has never been addressed: what is the impact of arbitrary nonlinearity in the interaction potential? Indeed, all studies were done with the cubic and quartic ones. A generic nonlinear interaction potential yields higher order nonlinear terms in the Taylor series expansion about equilibrium. As the magnitude of these terms increases with the energy it could be expected that these terms, and not the qubic and quartic, determine the delocalization and stability of QBs, and, broadly, thermalization properties of the systems.

In this Letter we present the theory of QBs in acoustic chains with arbitrary nonlinearity index in the interaction potential, exemplifying in the FPU chain. We demonstrate that there exists a critical nonlinearity index $\gamma^*=6$, such that for lower order nonlinearities the delocalization and stability thresholds of QBs approach zero as the chain length scales to infinity, while for higher orders they diverge. We find that the critical index value plays similar role in thermalization of the chain, which is rapid below it and non-observable above. It follows that only three lowest order nonlinear interaction terms are decisive for localization of energy or thermalization of generic acoustic chains.

The FPU system models an atomic chain by $N$ equal mass particles coupled by springs with linear and nonlinear interaction terms: 
\begin{equation}
\label{eq1}
\begin{aligned}
&\ddot{x}_n=(x_{n+1}-2x_n+x_{n-1})+\chi[(x_{n+1}-x_n)^{\gamma-1}\\
&-(x_n-x_{n-1})^{\gamma-1}],
\end{aligned}
\end{equation}
where $x_n$ is the deviation of the $n$-th particle from the equilibrium, and the boundary conditions are fixed $x_0=x_{N+1}=0$.
The classic $\alpha-$ and $\beta-$FPU models arise for $\gamma=3$, $\gamma=4$ respectively.

Canonical transformation
$x_n(t)=\sqrt{\frac{2}{N+1}}\sum\limits_{q=1}^N
Q_q(t)\sin{\left(\frac{\pi q n}{N+1}\right)}$ defines the reciprocal space of $N$ normal modes with the amplitudes standing for coordinates
$Q_q(t)$. Dynamical equations in this case read

\begin{equation}
\label{eq2}
\begin{aligned}
 &\Ddot{Q}_q+\omega_q^2 Q_q=-\frac{\chi\omega_q}{[2(N+1)]^{\gamma/2-1}} \sum\limits_{q_1,\ldots,q_{\gamma-1}=1}^N
 B_{q,q_1,\ldots,q_{\gamma-1}}\\
 &\prod\limits_{k=1}^{\gamma-1}\omega_{q_k} Q_{q_k},
 \end{aligned}
\end{equation}
where
$\omega_q=2\sin{(\pi q/2(N+1))}$ are the normal mode frequencies, and
\begin{equation}
\label{eq3}
 B_{q, q_1,\ldots,q_{\gamma-1}}=\sum\limits_{m=0}^{[\gamma/2-1]}\sum\limits_{\pm} (-1)^m \delta_{q\pm q_1 \pm \ldots \pm q_{\nu-1},2m(N+1)}
\end{equation}
are the intermode coupling coefficients, defining the long-range coupling between $q$-oscillators. 

In order to establish the continuation of QB in the nonlinear regime we set $\chi=0$, the energy $E_{q_0}=E$ in the normal mode $q=q_0$, while the other $q$-oscillators are at rest. These initial conditions correspond to the periodic trajectory in the phase space of the linear system. Following
\cite{mackayaubry}, the absence of resonances $n \omega_{q_0} \neq \omega_{q \neq q_0}$ is a sufficient condition for the continuation and the latter holds for generic finite $N$ \cite{oldstuff}.

Following this argument we develop the perturbation theory for (\ref{eq2}) in powers of the small parameter  $\sigma=\chi/[2(N+1)]^{\gamma/2-1}$: \ $\hat{Q}_q(t)=\sum_{i=0}^\infty\sigma^i Q_q^{(i)}(t)$. The frequency of the nonlinear mode is sought in the form $\hat{\omega}_q=\sum_{i=0}^\infty\sigma^i \omega_q^{(i)}(t)$ by means of the Lindstedt-Poincare method.
Take the low-frequency $q_0\ll N$ linear mode $Q_{q_0}(t)=A_0 \cos{\omega_{q_0}t}, \ Q_q(t)=0, \ q\neq q_0$ and its linear frequency $\omega_{q_0}^{(0)}=\omega_{q_0}$ as a zero-order approximation. The structure of the coupling coefficients determines that the only non-zero first order terms are $Q_{(\kappa-1)q_0}^{(1)}(t)$, where $\kappa=4,6,\ldots,\gamma$ for even  and $\kappa=3,5,\ldots,\gamma$ for odd $\gamma$. The coupling coefficients wrap up as $B_{(\kappa-1)q_0,q_0,\ldots,q_0}=C^{(\gamma-\kappa)/2}_{\gamma-1}$, and, finally, the equations of motion get the form of linear forced oscillators:

\begin{equation}
\label{eq6}
\begin{aligned}
 &\Ddot{Q}^{(1)}_{(\kappa-1)q_0}+\omega_{(\kappa-1)q_0}^2 Q^{(1)}_{(\kappa-1)q_0}=-\omega_{(\kappa-1)q_0} C^{(\gamma-\kappa)/2}_{\gamma-1} \\
 &\left(\omega_{q_0} A_0 \cos{\omega_{q_0}t} \right)^{\gamma-1}
 \end{aligned}
\end{equation}
The amplitude of the response on the resonant frequency $(\kappa-1)\omega_{q_0}$ reads
\begin{equation}
\label{eq7}
\begin{aligned}
& A_{(\kappa-1)q_0}=\sigma \frac{\omega_{(\kappa-1)q_0} \left(C^{(\gamma-\kappa)/2}_{\gamma-1}\right)^2 \omega_{q_0}^{\gamma-1} A_0^{\gamma-1}}{2^{\gamma-2}\left[(\kappa-1)^2\omega_{q_0}^2-\omega_{(\kappa-1)q_0}^2\right]} \\
 &\approx \sigma \frac{12 (N+1)^3 \left(C^{(\gamma-\kappa)/2}_{\gamma-1}\right)^2 \omega_{q_0}^{\gamma-1} A_0^{\gamma-1}}{2^{\gamma-2} \pi^3 q_0^3 (\kappa-1)\kappa (\kappa-2)},
 \end{aligned}
\end{equation}
and the frequency of the periodic orbit is given by
\begin{equation}
\label{eq7a}
\begin{aligned}
&\hat{\omega}_{q_0}=\omega_{q_0}\left[1+\frac{\chi}{2^{\gamma-1}}\left(C^{\gamma/2}_{\gamma-1}\right)^2\left(\frac{E}{N+1}\right)^{\gamma/2-1}\right],
 \end{aligned}
\end{equation}

Rewriting (\ref{eq7}) in terms of linear mode energies $E_q=\omega_q^2 A_q^2/2$, we obtain localization factors $\lambda_{\gamma,\kappa}$:
\begin{equation}
\label{eq10}
\begin{aligned}
&E_{(\kappa-1)q_0}=\lambda_{\gamma,\kappa}^2 E_{q_0},\\
&\lambda_{\gamma,\kappa}= \frac{3 \chi  \left(C^{(\gamma-\kappa)/2}_{\gamma-1}\right)^2 E_{q_0}^{\gamma/2-1}}{2^{\gamma-4} \pi^2 q_0^2 \kappa (\kappa-2)(N+1)^{\gamma/2-3}}
\end{aligned}
\end{equation}

One can also define the localization length in the mode space $\xi_{\gamma,\kappa}$ such that $\ E_{(\kappa-1)q_0}=e^{(\kappa-2)q_0/\xi_{\gamma,\kappa}} E_{q_0}$. Then it follows $\xi_{\gamma,\kappa}=-(\kappa-2)q_0/(2\ln{\lambda_{\gamma,\kappa}})$. 

Delocalization criterion is, thus, $\max\limits_\kappa\{\lambda_{\gamma,\kappa}\}\sim1$.  It corresponds to strong interaction between the linear modes and the breakup of perturbation theory. 

The crucial property of (\ref{eq10}) is the scaling of localization factor $\lambda$ with the system size. For $\gamma<6$ there is a delocalization threshold in $N$ above which the corresponding QB delocalizes. For $\gamma=6$ the localization is independent of the system size in the leading order approximation, while for $\gamma>6$ the localization strengths in longer chains. 

\begin{figure}[t]
{\centering
\resizebox*{\columnwidth}{!}{\includegraphics{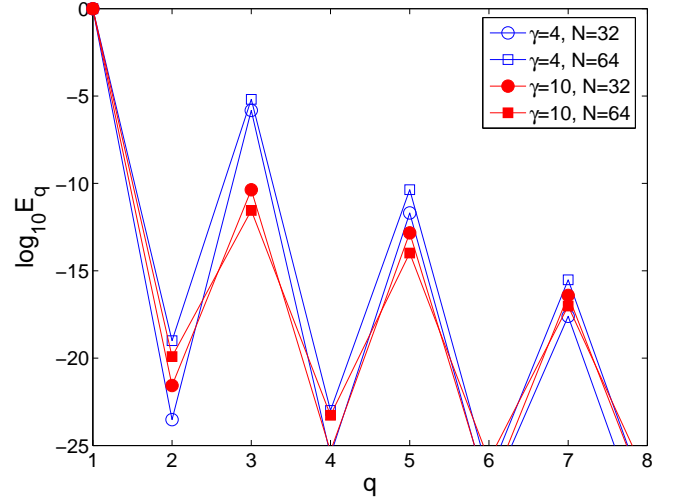}}}
{\caption{Localization of QBs in dependence on the chain size for the nonlinearity index below and above the critical $\gamma=6$. Here $q_0=1, \ \chi=0.001$, $\gamma=4$ and $\gamma=10$} \label{fig1}}
\end{figure}

Existence of the critical $\gamma^*=6$ is of principal importance for the general problem of thermalization and equipartition. As indicated above, the thermalization/strong chaos threshold in $\alpha$- and $\beta$-FPU is directly related to delocalization of QBs. It follows, that above the critical nonlinearity index the onset of strong chaos is retarded to much higher energies; moreover, the threshold in energy will grow with the system size. The second consequence is that for a generic nonlinear interaction potential the thermalization threshold will be fully determined by the three lowest-order nonlinear terms ($\gamma<6$) in the Taylor expansion about an equilibrium. 

Let us illustrate theoretical predictions with numerical results. First, we compute the QBs in (\ref{eq1}) for different nonlinearity index, following the numerical scheme developed in \cite{we_qb}. It is clearly seen that localization weakens for $\gamma<6$ and improves for $\gamma>6$ with the increase of the system size $N$ (Fig.\ref{fig1}). 

\begin{figure}[t]
{\centering
\resizebox*{\columnwidth}{!}{\includegraphics{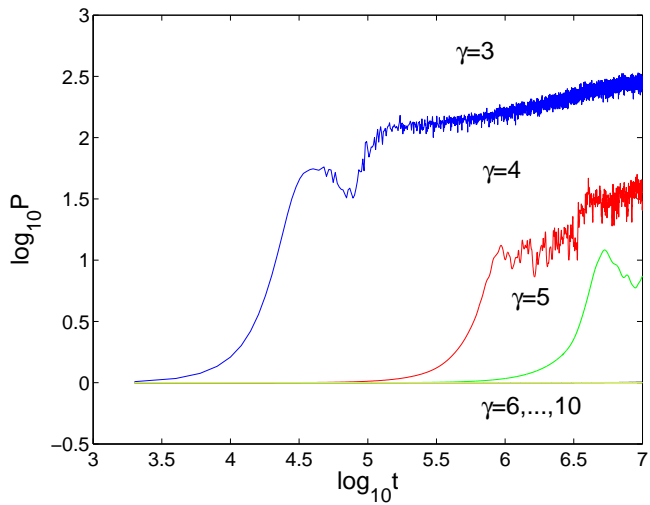}}}
{\caption{Evolution of participation number in dependence on the nonlinearity index. Here the mode $q_0=1$ is excited to $E=1$ at $t=0$, $\chi=1, \ N=1000$} \label{fig2}}
\end{figure}

To probe the dependence of the thermalization process on the nonlinearity index directly we perform the long-time integration of (\ref{eq1}) with the initial conditions corresponding to the lowest mode excited to the energy $E=1$. The other parameters are $N=1000, \ \chi=1$ and the integration time $t=10^7$ is of the order of $5000$ oscillation periods of the seed mode. To characterize the number of effectively excited modes we calculate the time dependence of the so-called participation number: $P=1/(\sum_{q=1}^N e_q^2)$, where $e_q=E_q/E$ are the normalized modal energies. (It is easy to see, that if, say, $m$ modes are excited to the same energy and the energy of the others is $0$ then $P=m$). This measure allows us to visualize the mode excitation process and the results clearly show thermalization for $\gamma<6$ and its absence (on the observed timescale) for the higher order nonlinearities (Fig.\ref{fig2}).

Let us study the linear stability of QBs for even nonlinearity indexes (note that instabilities are suppressed for odd $\gamma=3$ in the whole region of existence of the localized QB solution \cite{we_qb}). We keep in mind that it is tantamount to the onset of weak chaos, when chaotic trajectories stay localized in the mode space on the computationally accessible times, but are believed to thermalize through Arnold diffusion, eventually \cite{deLuca,we_qb}.   
For that we linearize equations of motion (\ref{eq2}) around the QB solution $Q_{q}=\hat{Q}_{q}(t)+\xi_{q}(t)$, centered at $q_0$ and having the frequency $\hat{\omega}_{q_0}$:
\begin{equation}
\label{eq11}
\begin{aligned}
 &\Ddot{\xi}_q+\omega_q^2 \xi_q=-\frac{\chi (\gamma-1)\omega_q}{[2(N+1)]^{\gamma/2-1}} \sum\limits_{q_1,\ldots,q_{\gamma-1}=1}^N
 B_{q_1,\ldots,q_n}\\
 &\left[\prod\limits_{k=2}^{\gamma-1}\omega_{q_k} \hat{Q}_{q_k}\right]\omega_{q_1}\xi_{q_1}
 \end{aligned}
\end{equation}
For $\gamma=4$ the strongest instability comes from primary parametric resonance in (\ref{eq11}) and involves pairs of resonant modes $\tilde{q},\tilde{p}=q_0\pm m, \ m=1, 2, 3, \ldots$. For higher indexes the number of potential instabilities increases and one needs to take into account all pairs of the type $\tilde{q},\tilde{p}=(\kappa-2)q_0\pm m, \ m=1, 2, 3, \ldots, \ \kappa=4, 6, \ldots, \gamma-2$. For each pair one excludes non-resonant terms and obtains 
\begin{equation}
\label{eq12}
\begin{aligned}
 &\Ddot{\xi}_{\tilde{q},\tilde{p}}+\omega_{\tilde{q},\tilde{p}}^2 \xi_{\tilde{q},\tilde{p}}=-\nu\omega_{\tilde{q}}\omega_{\tilde{p}}\xi_{\tilde{p},\tilde{q}}\cos[(\kappa-2)\hat{\omega}_{q_0}t]\\
 &-2\nu h \omega_{\tilde{q},\tilde{p}}^2\xi_{\tilde{q},\tilde{p}},  
 \end{aligned}
\end{equation}
where
\begin{equation}
\label{eq13}
\begin{aligned}
 &\nu=-\frac{\chi (\gamma-1)}{2^{\gamma-3}}\left(\frac{E}{N+1}\right)^{\gamma/2-1}\left(C^{(\gamma-\kappa)/2}_{\gamma-2}\right)^2,\\
 &h=\frac{1}{4}\left[C^{\gamma/2-1}_{\gamma-2}/C^{(\gamma-\kappa)/2}_{\gamma-2}\right]^2
 \end{aligned}
\end{equation}
Standard stability analysis \cite{we_qb} reveals that the first instability to happen corresponds to $\kappa=4$ and $m=1$ so that the instability condition reads:
\begin{equation}\label{eq14}
\left(C^{\gamma/2-2}_{\gamma-2}\right)^2 \frac{(\gamma-1)\chi}{2^{\gamma-5}\pi^2}\frac{E^{\gamma/2-1}}{(N+1)^{\gamma/2-3}}>1
\end{equation}

\begin{figure}[t]
{\centering
\resizebox*{\columnwidth}{!}{\includegraphics{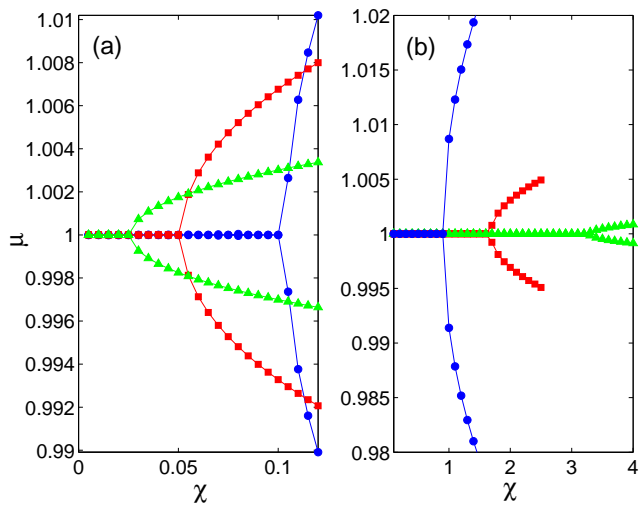}}}
{\caption{Dependence of the absolute values of the Floquet multipliers $\mu$ of QBs on the nonlinearity strength for (a) $\gamma=4$ and (b) $\gamma=8$. Circles mark $N=16$, squares -- $N=32$, triangles -- $N=64$} \label{fig3}}
\end{figure}

\begin{figure}[t]
{\centering
\resizebox*{\columnwidth}{!}{\includegraphics{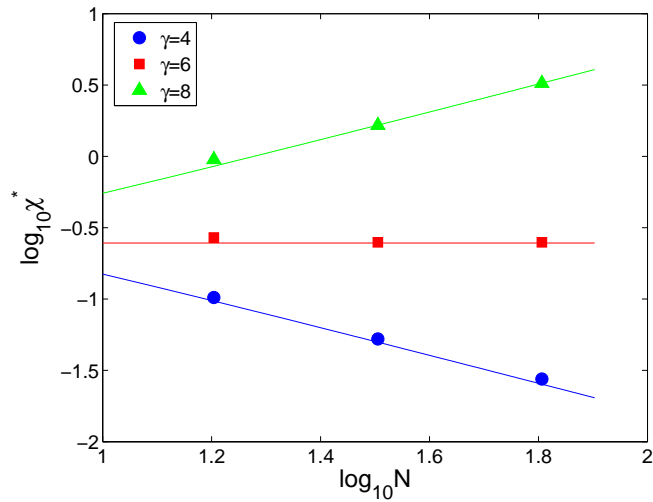}}}
{\caption{Theoretical (\ref{eq14}), solid lines, and numerical, markers, instability thresholds in dependence on the nonlinearity index $\gamma$} \label{fig4}}
\end{figure}

Remarkably, the same critical nonlinearity index $\gamma^*=6$ characterizes the scaling of the instability threshold $\chi^*$ with the system size $N$, as it does for localization. For $\gamma<6$ the instability threshold gradually diminishes as $N$ grows; for any fixed set of the energy and nonlinearity parameters there exist such $N$ above which all QBs will become unstable. On the contrary, for $\gamma>6$ the instability threshold grows with the system size; even if QBs were unstable in small systems they will attain stability if the chain length exceeds a threshold determined by (\ref{eq14}). Recalling the relation between instability and the onset of weak chaos, we come to similar conclusions as to the strong chaos threshold. Namely, for the FPU systems with the nonlinearity index higher than the critical one, weak chaos gets suppressed in large chains. In case of generic nonlinear coupling potential only three lowest order nonlinear terms in the expansion about the equilibrium influence and cause the development of weak chaos regime. 

Finally, we report numerical results on the QB stability. We compute Floquet multipliers for QBs centered at $q_0=3$ in dependence on the nonlinearity coefficient taking $E=1$, $N=16, 32, 64$, and $\gamma=4, 6, 8$. In accord with (\ref{eq14}), the instability threshold $\max\{\mu\}>1$ increases with $N$ for $\gamma=4$ and decreases for $\gamma=8$ (Fig.\ref{fig3}). The quantitative agreement between the theoretical predictions for $\chi^*$ and numerics is very good (Fig.\ref{fig4}).

In conclusion, we develop the theory of q-breathers in acoustic chains with arbitrary nonlinearity index in the coupling potential. Theoretical and numerical studies of the mode space localization and stability reveal that there exists the critical nonlinearity index $\gamma^*=6$, below which q-breathers lose localization and stability as the system size is increased. In a marked contrast, these properties strengthen in larger chains for higher order nonlinearities. According to the direct relation between delocalization/instability and strong/weak chaos thresholds, the latter must demonstrate the same regularities. Indeed, the long-time evolution from the lowest mode excitation show fast thermalization below the critical nonlinearity index and its absence above. For a generic interaction potential, localization of energy and thermalization processes are determined only by three lowest order nonlinear terms in its power series expansion; if they are absent, then thermalization is strongly suppressed in large systems and extremely high energies are required for it to be observed. 

Beside the fundamental physical problems of thermalization and ergodicity, the results build on the theory of nonlinear oscillations and energy localization in spatially discrete systems with the high demand coming from experiments and applications with micro- and nano-scale electro-mechanical oscillator arrays \cite{Roukes1, Sievers, Roukes2}.

I thank S. Flach, O.~I.~Kanakov and K.~G.~Mishagin for useful discussions.

\end{document}